\documentclass{emulateapj}

\usepackage{epsfig}

\makeatletter
\newenvironment{inlinefigure}{%
\def\@captype{figure}%
\noindent\begin{minipage}{0.999\linewidth}\begin{center}}
{\end{center}\end{minipage}\smallskip}
\makeatother

\newcommand{\etal}{\mbox{et al.}}

\newcommand{\msun}{{$M_{\odot}$}}

\newcommand{\ergsec}{erg s$^{-1}$}
\newcommand{\rxte}{{\it RXTE}}
\newcommand{\sax}{{\it BeppoSAX}}
\newcommand{\asca}{{\it ASCA}}
\newcommand{\exosat}{{\it EXOSAT}}

\newcommand{\degrees}{$^{\circ}$}

\newcommand{\tertwo}{1E 1724--307}

\newcommand{\thankhubble}{M. Muno was supported through a Hubble Fellowship grant (program number HST-HF-01164.01-A) from the Space Telescope Science Institute, which is operated by the Association of Universities for Research in Astronomy, Incorportated, Under NASA contract NAS5-26555.}

\shortauthors{Muno \etal}
\shorttitle{Radio Emission and Hard X-rays}

\begin{document}

\title{A Lack of Radio Emission from Neutron Star Low Mass X-ray Binaries}
\author{Michael P. Muno,\altaffilmark{1,2} 
Tomaso Belloni,\altaffilmark{3}
Vivek Dhawan,\altaffilmark{4}
Edward H. Morgan,\altaffilmark{5} Ronald A. Remillard,\altaffilmark{5}
and Michael P. Rupen\altaffilmark{4}}

\altaffiltext{1}{Department of Physics and Astronomy, University of California,
Los Angeles, CA 90095; mmuno@astro.ucla.edu}
\altaffiltext{2}{Hubble Fellow}
\altaffiltext{3}{INAF --- Osservatorio Astronomico di Brera, Via E. Bianchi 46, I-23807, Merate, Italy}
\altaffiltext{4}{National Radio Astronomy Observatory, Soccorro, NM 87801}
\altaffiltext{5}{Center for Space Research,
Massachusetts Institute of Technology, Cambridge, MA 02139}

\begin{abstract}
We report strict upper limits to the radio luminosities of three
neutron star low-mass X-ray binaries obtained with the Very Large Array while 
they were in hard X-ray states as observed with the {\it Rossi X-ray Timing
Explorer}:
\tertwo, 4U 1812--12, and SLX 1735--269. We compare these upper limits to 
the radio luminosities of several black hole binaries in very similar hard 
states, and find that the neutron star systems are as faint as or 
fainter than all of the black hole candidates. The differences in luminosities 
can partly be attributed to the lower masses of the neutron star systems, 
which on theoretical and observational grounds are expected to decrease the 
radio luminosities as $M^{0.8}$. However, there still remains a factor of 30 
scatter in the radio luminosities of black hole and neutron star X-ray 
binaries, particularly at X-ray luminosities of a few percent Eddington. We 
find no obvious differences 
in the X-ray timing and spectral properties that can be correlated with the 
radio luminosity. We discuss the implications of these results on current 
models for the relationship between accretion and jets.
\end{abstract}


\section{Introduction}

Astrophysical jets are observed from a wide range of systems, including 
active galactic nuclei \citep{lb96}, 
young stellar objects \citep{bac96}, accreting white
dwarfs in symbiotic binaries \citep{cro02,sk03} and super-soft X-ray 
sources \citep{cow98}, and neutron stars and
black holes in X-ray binaries \citep{fen04}. Although 
it is generally accepted that the power supplied from these jets has its 
origin in accretion, the details of the mechanisms producing the jets are 
not fully established 
\citep[e.g.,][Falcke, K\"{o}rding, \& Markoff 2004]{bp82,mku01,hs03}. 
Observationally, one possible impediment to understanding these 
mechanisms is that the jets probably form on the time scales at which
changes occur in the accretion flow 
\citep[e.g., the viscous time scale; see][]{hk96,mir98,sk03}. 
For young stellar objects and active galactic nuclei, that time scale is 
months to years, which makes it difficult to obtain well-sampled observations
of jets as they form. Therefore, in trying to establish what conditions are
needed to produce jets, it is often hard to disentangle the relative 
importance of the state of the accretion flow from the physical properties 
of the systems as a whole, such as the binary orbital period or the rotation 
rate of the accreting object. 
The exceptions are accreting compact objects, in which, because of their 
small size, the structure of their accretion flows are observed to change 
on time 
scales of hours to days. Recent observations of X-ray binaries have provided 
the first information about the time-dependent relationship between the 
accretion flow and the formation of jets (e.g., Mirabel \etal\ 1998;
Fomalont, Geldzahler, \& Bradshaw 2001; Fender \etal\ 2004; Fender, Gallo,
\& Belloni 2004).

A particularly striking picture of the relationship between accretion and 
jets has emerged for black hole X-ray binaries 
\citep[see][for a review]{fen04}. The accretion flows can 
be characterized by three ``states'' based on the spectral and timing 
properties of their X-ray and gamma-ray emission (0.5--500 keV): 
(1) a hard state that usually occurs at relatively low accretion rates in 
which the X-ray 
emission forms a power-law with photon index $\Gamma \approx 1.5-2.0$ that is  
exponentially cutoff at an energy of $\sim 100$~keV, and the power spectrum
resembles band-limited white noise with a power of 10--30\% rms, (2) 
a thermal state at higher accretion rates, in 
which the X-ray emission resembles that expected from a canonical 
optically-thick, geometrically-thin accretion disk with a temperature of 
$\sim 1$~keV \citep{ss73}, and in which the power spectrum exhibits only
weak noise (1-6\% rms) with a power-law shape; and (3) a 
steep power-law
state that is characterized by the sum of thermal emission plus a power-law
of photon index $\Gamma = 2 - 3$ that extends without a break to 
$\sim 500$ keV, and the power spectrum usually contains quasi-periodic
oscillations at low-frequency (0.1--20~Hz), and less often at high 
frequencies (150--450~Hz) or in the mHz range
\citep[see][for further discussion]{mr04}. 

The radio emission
associated with these states comes in two forms. First, optically thick radio 
emission from compact jets that extend only tens of AU 
\citep[Dhawan, Mirabel, \& Rodr\'{\i}guez 2000;][]{sti01}
is observed to coincide with the X-ray hard power-law 
state \citep[$\Gamma = 1.5-2.0$;][Gallo, Fender, \& Pooley 2003]{bro99,cor00}.
The hard states and their jets can persist for days to months with 
relatively steady flux, and yet the jets become quenched when sources enter 
thermal-dominated states.\footnote{Note that it is unclear whether optically
thick radio emission accompanies the rarer steep power-law states
\citep[e.g.,][]{tav96, fbg04}.} Second, transitions 
between these states are often observed to coincide with the formation 
of discrete synchrotron-emitting structures that travel relativistically
across the sky \citep[e.g.,][]{mr94,fen99b,fgb01,fk01}. 
These ballistic jets form on time scales of a day, and eventually travel 
thousands of AU from the 
central black hole. 

X-ray binaries containing neutron stars with relatively weak magnetic fields
($B \la 10^9$ Gauss; these are almost always found in low-mass X-ray binaries,
or LMXBs) resemble black hole systems in some respects, because 
the inner edge of the accretion disk can extend to near the inner most
stable General Relativistic orbit in both types of systems. However, the 
relationship between X-ray and radio emission from neutron star LMXBs 
has not been as firmly established. Similarly to the black hole X-ray 
binaries, neutron star LMXBs exhibit a hard X-ray state with a 
$\Gamma \approx 1.5-2.0$. However, the analogs of the thermal-dominated 
and steep power-law states appear to be replaced by soft states in which 
most of the X-ray emission is produced by the boundary between the accretion 
disk and the neutron star \citep[e.g.,][]{vdk95,dg03}. 
Transient radio emission 
has been observed from neutron star X-ray binaries during periods in which 
their X-ray spectrum was highly variable 
\citep[e.g.,]{pen88,ber00,hom03,mig03}, and in two cases this emission
was resolved into ballistic jets \citep[][]{fgb01,fen04b}.
However, the peak luminosity 
of their radio emission is a factor of $\approx 30$ lower than the black hole 
systems \citet{fk01}.Two neutron star system were also detected 
in persistently soft states \citep[4U 1820--30 and Ser X-1;][]{mig04}; 
this emission does not easily fit into the paradigm derived from black holes.
However, no study has systematically targeted neutron star X-ray binaries 
in their hard, $\Gamma = 1.5-2.0$ power-law states to determine whether 
or not they exhibit steady, compact radio jets.

In this paper, we search for radio emission from three neutron star 
LMXBs in hard X-ray states with the Very Large Array (VLA) 
in order to determine whether the formation of compact jets affected by
changing the central compact object from a black hole to a neutron star.
We have observed three systems that remained in hard, $\Gamma = 1.5-2.0$ 
power-law states for the first five years of observations with
the {\it Rossi X-ray Timing Explorer} (\rxte):
\tertwo\ in Terzan 2, SLX 1735--269, and 4U 1812--12.
Each of these systems exhibits thermonuclear X-ray bursts, which demonstrates
unambiguously that they contain neutron stars 
\citep{gri80,mur83,baz97}. However, their X-ray properties are 
otherwise almost indistinguishable from those of black holes in their hard 
states \citep[][]{oli98,wvk99,bar99,bar00,bel02,bar03}. 

The organization of this paper is as follows. In Section~\ref{sec:xte}, we
explain how we selected our sample of sources, and 
report on the X-ray emission derived with \rxte. 
We include considerable detail in order to establish the 
degree to which these systems resemble black hole X-ray binaries in the 
hard state. 
In Section~\ref{sec:vla}, we 
report on strict upper limits to the radio emission from three LMXBs in our
sample that were obtained with the VLA, nearly simultaneous with the \rxte\
observations.
In Section~3, we discuss the 
implications of a lack of radio emission from neutron star LMXBs in the hard
state.

\section{Observations}

The basic observational properties of the three neutron star LMXBs that we have
studied are listed in Tables \ref{tab:vla} and \ref{tab:xte}.

\subsection{\rxte\label{sec:xte}}

\rxte\ carries three instruments: the All-Sky Monitor (ASM), consisting
of three, 30 cm$^{2}$ proportional counters sensitive between 1.5--12 keV
that scan 80\% of the sky every 90 minutes \citep{lev96}; 
the Proportional Counter Array (PCA), consisting of five, 1300 cm$^{2}$ 
detectors that record events with capability for microsecond time resolution 
and 256 channels of spectral resolution between 2--60 keV 
\citep{jah96}; and the High Energy X-ray Timing Experiment, consisting
of two clusters of scintillation detectors, each with an effective area of
800 cm$^2$ and sensitivity between 15--250 keV \citep[HEXTE;][]{roth98}.
Data from all three instruments was used to characterize the X-ray emission 
from the LMXBs in our sample.

We first used data from the PCA and ASM to identify several neutron star 
LMXBs that persistently remained in a hard state (sometimes also referred 
to as the ``island'' of the atoll-shape in the color-color diagram) from 
the sample of systems that exhibit thermonuclear X-ray bursts 
(D. Galloway \etal, in preparation). 
We first confirmed
that the X-ray emission was steady using data from the \rxte\ ASM.
We then measured the hardness of 
the X-ray spectrum using a hard color, which was defined as the ratio of the 
background-subtracted PCA counts in the (8.6--18.0)/(5.0--8.6) keV 
energy bands, where the counts were normalized to account for changes
in the gain of the proportional counter units \citep[see][]{mrc03}. 
%
Eight sources qualified as persistently hard,
and we were able to obtain \rxte\ and VLA observations with a day of each 
other for three of these sources: 4U 1812--12, SLX 1735--269, and 
\tertwo.\footnote{ 
The sample of persistently hard neutron star LMXBs also includes 
4U 1323--619, SAX J1712.6--3739, GS 1826--25, 4U 1832--330, and 4U 1850--087.}
The ASM light curves and PCA color-color diagrams from 8 years 
of observations are displayed 
\begin{figure*}[th]
\centerline{\epsfig{file=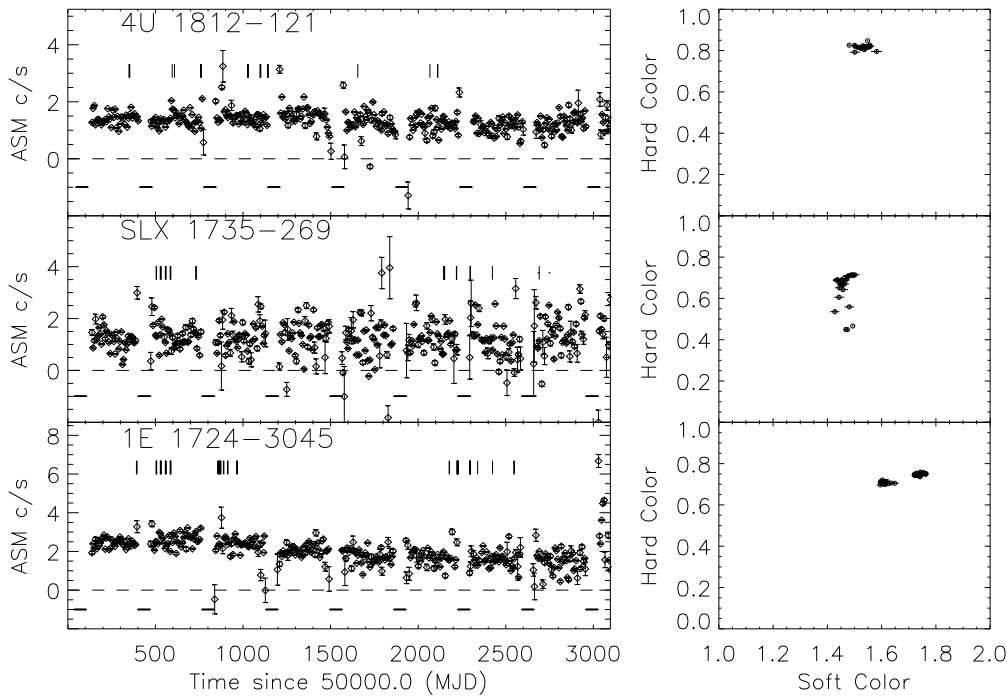,width=0.8\linewidth}}
\caption{ASM light curves and PCA color-color diagrams illustrating the 
stability of the X-ray emission from the three sources we studied. The 
{\it left panels} plot weekly averages of the individual 90 s dwells. The
horizontal bars at a count rate of $-1$ indicate periods when the sun was 
within 35\degrees\ of the source. The vertical bars indicate the times of 
PCA observations, a subset of which were analyzed in detail for this paper.
The {\it right panels} contain the hard and soft colors from the PCA
data, which are defined as the ratio of the background-subtracted 
detector counts in the (3.6--5.0)/(2.2--3.6) keV and the 
(8.6--18.0)/(5.0--8.6) keV energy bands, respectively. All of the sources
remained in the hard portion of the color-color diagram prior to 2001, 
although SLX 1735--269 became softer during observations in 2002 January.}
\label{fig:asm}
\end{figure*}
in Figure~\ref{fig:asm}.
In the remainder of this section, we report the X-ray spectral
and timing properties from \rxte\ observations taken within 6 weeks of 
the VLA observations.


\subsubsection{X-ray Spectra}

\begin{figure*}[th]
\centerline{\epsfig{file=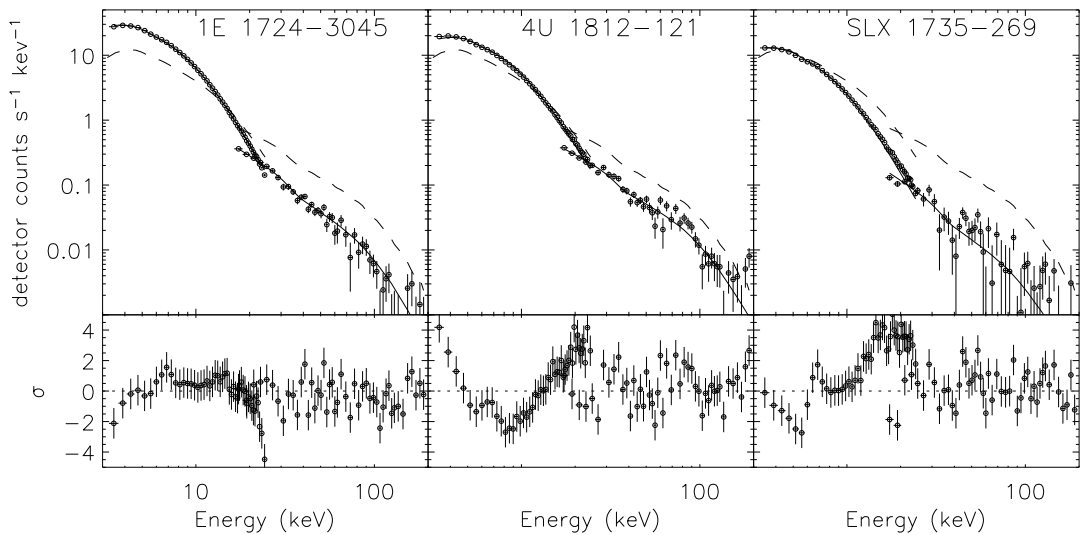,width=0.9\linewidth}}
\caption{PCA and HEXTE spectra of the sources in this study, in 
units of detector
counts. The residuals from the best-fit absorbed Comptonization spectra
are displayed in the bottom panels, in units of the uncertainty in the 
data. Residuals between 6--20 keV in 4U 1812--12 and SLX 1735--269 
are probably produced by hard photons that reflect off the optically
thick accretion disk. We have plotted the spectrum of Cyg X-1 in the 
hard state with a dashed line, for comparison to this sample of 
neutron star LMXBs.}
\label{fig:spec}
\end{figure*}

To produce spectra, we extracted data in 128 energy channels from the top 
layer of each active detector of the PCA, and in 64 channels from 
cluster 0 of HEXTE. For the PCA, the detector response and background were
estimated using standard tools in FTOOLS version 5.1. For HEXTE, the
background from the intervals in which the cluster was pointed off-source,
and the response and effective area estimates were obtained from 
CALDB version 2.23. We modeled the spectra jointly in XSPEC version 11.2, 
using the 3--25 keV energy range from the PCA, and the 15--200 keV energy
range from HEXTE.
We applied a constant normalization to account for differences in the 
calibrated effective area of the PCA and HEXTE, and a 1\% systematic 
uncertainty added in quadrature to the count spectra from the PCA to 
account for uncertainties in the detector calibration near the Xe edge
at 4.5 keV. After confirming that the spectrum did not vary significantly
over the course of the observations that we considered, we averaged the 
spectra to obtain better signal-to-noise at high energies. The resulting 
spectra are displayed in Figure~\ref{fig:spec}.

We modeled the spectra using a phenomenological model that included a 
power-law continuum that produced most of the 2.5--200 keV flux; 
a blackbody component at low energies to remove residuals below 4~keV,
which could result from emission from the neutron star's surface or the
inner accretion disk; a reflection component to account for residuals 
between 6--7 and 10--20 keV \citep[{\tt refsch} in {\tt XSPEC}][]{fab89,mz95}, 
which may be produced by hard X-rays impinging on an optically thick accretion
disk; and low-energy absorption caused by interstellar gas. 
The column density of interstellar matter ($N_{\rm H}$) was fixed to
the values determined from previous observations of each source by
either \asca\ or \sax\ \citep{dav97,gui98,bar99,bar03}. 
The parameters of the reflection component could not all be constrained 
independently, so we fixed the inner and outer radii of the disk to 10 
and 100 gravitational radii ($GM/c^2$), 
the power law index for reflection emissivity to -2, the metal abundances to
their solar values, the inclination to 30\degrees, the disk temperature
to 30,000 K, and the ionization fraction to 100. 
%
The final model is listed in Table~\ref{tab:spec}. The black body and 
reflection components were only needed to obtain an acceptable 
$\chi^2/\nu$ for SLX 1735--269 and 4U 1812--12, and so were not included
in the model of \tertwo.

The spectra of these neutron star LMXBs are 
\begin{figure*}[th]
\centerline{\epsfig{file=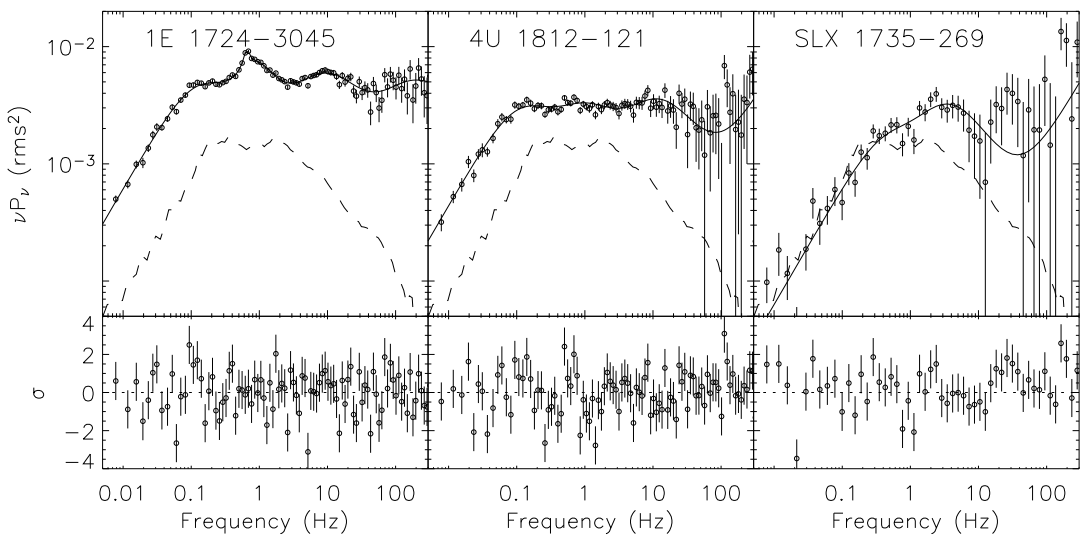,width=0.9\linewidth}}
\caption{Power spectra of the neutron star LMXBs. The {\it bottom panels}
illustrate the residuals to the fit of several Lorentzians to the data. 
The power spectrum of Cyg X-1 in the hard state is illustrated with 
the dashed line, which has been scaled downward by a factor of 10 for 
clarity. The timing data from the neutron star LMXBs have more power above 
20~Hz, but are otherwise very similar to that of Cyg X-1.}
\label{fig:pds}
\end{figure*}
qualitatively similar to those 
of black hole candidates in their hard states, such as 
Cyg X-1 \citep[e.g.,]{dis01}
and GX 339-4 \citep[e.g.]{wil99}.
To highlight this similarity, we plot the spectrum
of Cyg X-1 from a 21 ks \rxte\ exposure taken on 1996 October 23 in each of 
the panels in Figure~\ref{fig:spec}. Both the black holes and the neutron 
stars are dominated by Comptonized spectra of $kT \sim 100$ keV and
$\tau \la 1$, plus a cool $kT \la 1$ keV soft excess and reflection from 
an optically thick accretion disk. The spectra of the black hole candidates,
however, are often a bit flatter ($\Gamma \approx 1.5$) than the 
neutron star systems ($\Gamma \approx 2$).

\subsubsection{X-ray Timing}

We produced power spectra using PCA data with 122 $\mu$s ($2^{-13}$ s) time
resolution in a single energy channel (effectively 2--30~keV).
We computed power spectra in 256~s intervals, and averaged them weighted by 
the total counts to produce an accurate estimate of the continuum power.
We then subtracted the deadtime-corrected Poisson noise level from
the power spectra \citep{zha96}, re-binned 
\begin{figure*}[th]
\centerline{\epsfig{file=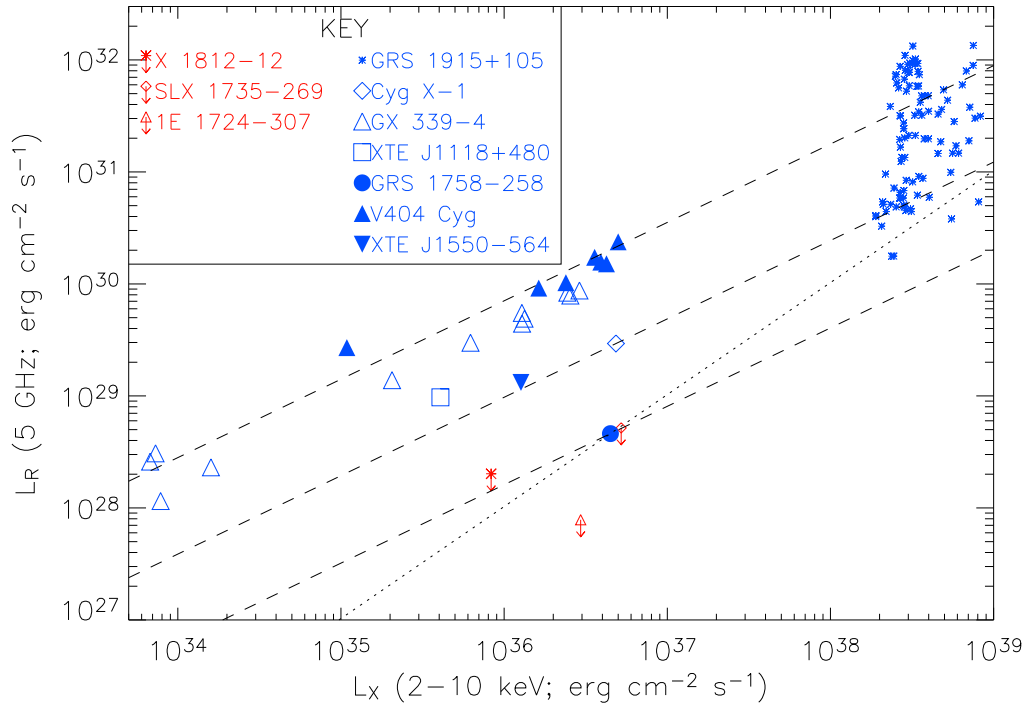,width=0.8\linewidth}}
\caption{Comparison of the radio and X-ray luminosity of the neutron 
star LMXBs in our sample (red symbols), with those from several black hole 
candidates in the hard X-ray state (blue symbols). 
The {\it dashed lines} represent the observed 
$L_{\rm R} \propto L_{\rm X}^{0.7}$ correlations observed for
several individual sources. The {\it dotted line} illustrates 
a linear $L_{\rm R} \propto L_{\rm X}$ proportionality that is
normalized to the radio flux from GRS 1758--258.
The radio flux is 
defined as that in a 5 GHz band around a 
central frequency of 5 GHz, assuming a flat radio spectrum. If 5 GHz measurements
were not available, then either measurements at other frequencies were 
interpolated onto 5 GHz, or measurements at a signle frequency were used under
the assumption that the spectrum was flat (this introduced at most a 10\% 
error). The X-ray luminosity is defined as that between 2--10 keV, deabsorbed. 
Values taken in different energy bands were extrapolated into the 2--10 keV band
using the observed spectrum, or in one case (V404 Cyg) assuming a 
$\Gamma = 1.5$ power law spectrum. Distances to the neutron star systems were 
obtained as follows: for \tertwo, we used the distance to Terzan 2 (4.1 kpc; 
Barbury, Bica, \& Ortolani 1998); for 4U 1812-12, we used the peak flux of an 
Eddington-limited burst 
\citep[6.6 kpc, taking $L_{\rm Edd}=3.5\times10^{38}$ erg s$^{-1}$ for pure He;][]{mur83};
for SLX 1735-269, we assumed that a burst that had no spectral information was
Eddington-limited and derived an upper limit to the distance 
\citep[$< 12$ kpc;][]{baz97}. 
For the black hole candidates, the distances
were taken from \citet{mr04}, except for GRS 1758-258, which was taken from
\citet{gfp03}.  
References for radio and X-ray fluxes of the black hole systems are as follows:  
GRS~1915$+$105 \citep{mun01}, Cyg~X-1 \citep{sti01,dis01}, 
GX~339$-$4 \citep{cor03}, XTE~J1118$+$480 \citep{mcc01,fen01}, 
V404 Cyg \citep{hh92}, XTE J1550-546 \citep{cor01,tom01}, 
GRS~1758$-$258 \citep{lin00}.}
\label{fig:lrlx}
\end{figure*}
them logarithmically, and 
estimated uncertainties from the standard deviation in the individual points
averaged to compute the final power spectrum.
To quantify the shape of the power spectrum, we 
modeled the continuum with several zero-centered Lorentzian functions, and 
any QPOs with Lorentzians with variable centroid frequencies. We added 
Lorentzians until
doing so no longer led to a significant decrease in $\chi^2$. The results are 
listed in Table~\ref{tab:pds}, and displayed in Figure~\ref{fig:pds}. 

As before, the power spectra of these LMXBs strongly resemble those from 
black hole candidates such as Cyg X-1 and GX 339-4 in the hard state
\citep{now99, sl99, now00}.
For reference, we have displayed the power spectrum of Cyg X-1 from 
1996 October 23 in each panel of Figure~\ref{fig:pds}. As noted by 
\citet{sr00} and Belloni \etal\ (2002)\nocite{bel02}, 
the main difference is that the neutron star systems 
exhibit more power continuum above 20 Hz.

\subsection{VLA\label{sec:vla}}

The VLA is a multi-frequency, multi-configuration aperture
synthesis imaging instrument, consisting of 27 antennas of 25 m
diameter.   
We obtained VLA observations of 4U 1812--12 under program 
AR 458, and of \tertwo\ and SLX 1735--269
under program AM 703 (Table~\ref{tab:vla}).
The observation under AR 458 was a 15 minute integration at 8.45 GHz.
The observations under program AM 703 were 1 hour integrations at 
1.42 GHz and 2 hour integrations at 5.0 and 8.45 GHz.
In all cases, we obtained 2 adjacent 
bands of 50 MHz nominal bandwidth processed in continuum mode.
%
Calibration and imaging were carried out with standard tasks in the
NRAO Astronomical Image Processing System (AIPS) package. 

None of the three sources were detected in the radio. We estimated 
1-$\sigma$ upper limits to the radio flux by measuring the rms dispersion 
in the noise within 5\arcsec\ of each source. These upper limits are 
listed in Table~\ref{tab:vla}.

\section{Discussion\label{sec:why}}

We have placed the first strict upper limits on the radio emission from 
neutron star LMXBs that are known to have been in hard X-ray states. 
In Figure ~\ref{fig:lrlx}, we compare the luminosities of these neutron 
star LMXBs in the radio and X-ray bands with those of several black hole
X-ray binaries. For the black holes, the luminosities were obtained from 
the references tabulated by 
\citet[][see the figure caption for details]{gfp03},
using the most current distances listed in \citet{mr04}. The neutron star
LMXBs in our sample have the lowest radio luminosities of the X-ray binaries
in Figure~\ref{fig:lrlx}.

Several authors have found that the relationship between the X-ray and radio
luminosities of both 
individual black hole systems and the ensemble of
systems follows the relationship $L_{\rm R} \propto L_{\rm X}^{0.7}$ 
\citep{cor03,gfp03}. Remarkably, this scaling follows the spectral energy 
distribution expected from models of a standard conical radio jet 
that extracts a fixed fraction of the total
\begin{inlinefigure}
\centerline{\epsfig{file=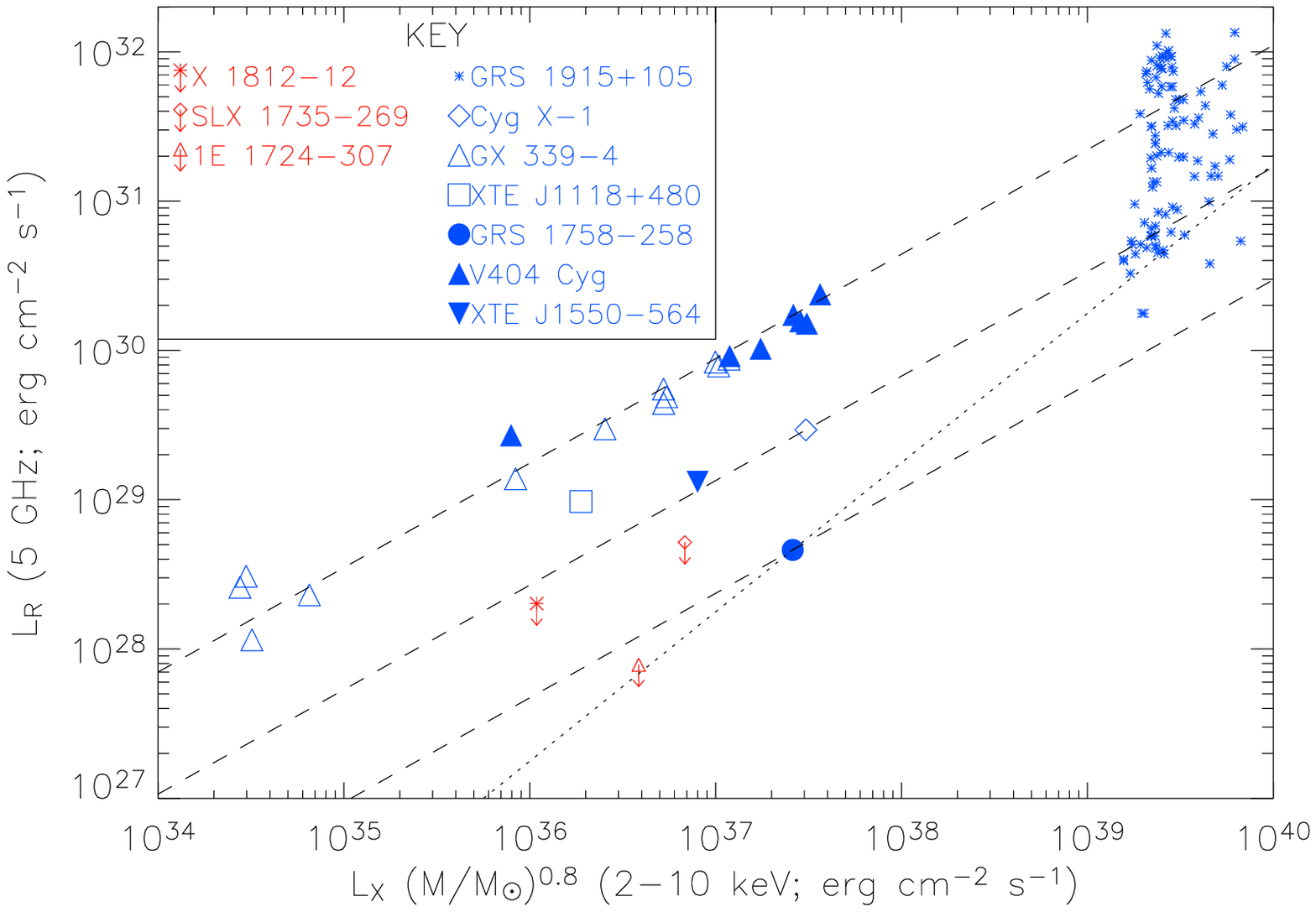,width=\linewidth}}
\caption{Same as Figure~4, except that the X-ray luminosity includes 
an extra mass-dependent term $(M/M_{\odot})^{0.8}$. Including this 
mass-dependence demonstrates that the radio upper limits from the three 
neutron star systems are all consistent with the fundamental plane 
defined by the faintest black hole system, GRS 1758--258.}
\label{fig:lrlx_m}
\end{inlinefigure}

\noindent
 energy from accretion 
\citep[][Merloni, Heinz, \& di Matteo 2003]{bk79,mar03,hs03}.
Therefore, we also plot in 
Figure~\ref{fig:lrlx} lines of $L_{\rm R} \propto L_{\rm X}^{0.7}$ that 
intercept data from V404 Cyg, Cyg X-1, and GRS 1758--258; these lines bound
the observed scatter in radio luminosities. 
The upper limits on the radio luminosities of the neutron star X-ray binaries
are equal to or below the line of $L_{\rm R} \propto L_{\rm X}^{0.7}$ that
passes through the faintest black hole X-ray binary observed in
the hard X-ray state, GRS 1758--258 \citep{lin00,mar02}. 
The strictest 1-$\sigma$ 
upper limit on the radio luminosity, from \tertwo, is a factor of $\approx 4$ 
below the relationship for GRS 1758--258. For comparison, there is a factor of 
$\approx 30$ difference in the radio luminosities of GRS 1758--258 and those
of V404 Cyg and GX 339--4. 

By comparing a sample of accreting black holes
in X-ray binaries and active galactic nucleii, \citet{mhm03} and
\citep{fkm04} have found
an additional dependence on black hole mass: 
$\log L_{\rm r} = 0.6 \log L_{\rm X} + 0.8 \log M$. This 
dependence is also expected from conical jet models
\citep{fkm04,rl04}. Therefore, in Figure~\ref{fig:lrlx_m} we re-plot
the data in Figure~\ref{fig:lrlx} with the $x$-axis scaled by a factor
$(M/M_{\odot})^{0.8}$.
Since all of the black holes in the sample have masses that are between
5--15 \msun\ \citep{mr04}, the proposed dependence on mass does not account 
for the scatter in the $L_{\rm R}$ vs. $L_{\rm X}$ relationship for
black holes. However, 
accounting for the fact that the neutron stars are likely to have masses 
near 1.4 \msun\ \citep{tc99} shifts them so that the upper limits on 
their radio luminosities are consistent with the relationship defined 
by GRS 1758--258. However, the mean luminosity of the neutron star LMXBs 
is still lower than that of the black holes.

Unfortunately, the individual upper limits on the radio emission from 
the three neutron star LMXBs are not strict enough to establish a clear 
difference between the production of jets from accreting black holes and 
neutron stars. Nevertheless, 
the low radio luminosities of the neutron star systems do raise further 
questions as to the origin of the dispersion in the radio luminosities of 
accreting X-ray binaries in hard X-ray states. \citet{gfp03} suggested that
GRS 1758--258 has an anomalously low radio luminosity because its jets have
been ``quenched'' by an unknown mechanism that begins to operate when black 
holes accrete at a few percent of the Eddington rate. Indeed, all four
sources with low radio luminosities have 
$L_{\rm X} \approx (0.01-0.04)L_{\rm Edd}$.
However, the quenching mechanism would have to activate in these systems 
without manifesting any obvious cause (or reaction)
in the energy spectrum and timing properties of the X-ray emission (see Lin 
\etal\ 2000 for GRS 1758--258, and Figures~\ref{fig:spec} and \ref{fig:pds} 
for the neutron star LMXBs). 

Further insight into the possible causes of the relative faintness of the 
the radio emission during the hard X-ray states of GRS 1758--258, \tertwo, 
4U 1812--12, and SLX 1735--269 can be found by considering 
current models explaining why hard X-ray and radio emission are observed 
together. For instance, several authors have suggested that the 
structure of the accretion flow is more conducive to launching jets in the 
hard state than in the soft, most likely because the geometry of the
accretion flow differs (Meier, Koide, \& Uchida 2001; Meier 2001;
Merloni \& Fabian 2002).
Specifically, the hard X-ray state is thought to occur when the 
accretion flow is radiatively inefficient and geometrically thick, whereas the 
thermal-dominate state is assumed to occur when the accretion disk is 
radiatively-efficient and 
geometrically thin (e.g., Esin, McClintock, \& Narayan 1997). At the 
same time, the jets are assumed to be formed by threading matter onto a 
poloidal magnetic field anchored to the rotating accretion flow 
\citep{bp82}, while the fields are generated by turbulent motions in the 
accretion flow. Since the scale heights of the turbulent motions are much 
larger when the flow is geometrically thick, stronger magnetic fields 
and more powerful jets are produced in the hard state. Under this model, 
there is no obvious reason to associate the faintness of the radio emission 
from GRS 1758--258 and the neutron star LMXBs with the ``quenching'' of what
would otherwise be a powerful compact jet, since the X-ray emission from 
these sources are almost indistinguishable from those of radio-luminous 
sources such as Cyg X-1. Instead, it seems more natural to assume that 
another physical or observational property of the radio-faint sources 
differs from the radio-luminous ones. 

Another basic explanation for the association is that the hard X-rays are 
produced by optically-thin synchrotron emission from the electrons in the jet 
(Markoff, Falcke, \& Fender 2001; Markoff \etal\ 2003; note that in this 
case the mechanism for launching the jet is unspecified). Under this model,
there is little freedom to assume that the jet power is simply lower in 
GRS 1758-298 and the neutron star LMXBs during their hard states, because 
the X-ray emission is produced in the jet. One option is to assume that 
the break frequency at which the synchrotron emission from the jet changes 
from optically thick to optically thin shifts to higher frequencies, so 
that the X-ray luminosity from the jet is preserved, while the radio 
luminosity at 5 GHz declines. This could be accommodated if the shock 
that accelerates the X-ray 
emitting electrons in the jet forms closer to the accreting compact object. 
Since the shock is positioned $\sim 10^3$ gravitational radii from the black 
holes in the models 
for XTE J1118+480 and GX 339--4 presented by \citet{mff01,mar03}, in principle
there is plenty of room for it to move. Moreover, the presence of stronger
noise in the power spectra of the neutron star LMXBs above $\sim 20$~Hz
in Figure~\ref{fig:pds} could also be taken as evidence that the X-ray
emitting region is closer to the compact object.\footnote{It has not 
been reported in the literature whether high frequency noise is stronger 
in GRS 1758--258 than in, e.g., Cyg X-1 and GX 339--4.} Alternatively, 
some of the X-ray emission may be produced by synchrotron self-Compton
emission, which would allow for some decoupling between the X-ray and 
radio emission (S. Markoff, M. Nowak, \& J. Wilms, in preparation).

Given the tentativeness of the above arguments, we believe it also is 
important to consider whether
the dispersion observed in the radio luminosity from the ensemble
of black hole and neutron star X-ray binaries is caused by differences in 
the physical or observational parameters of the systems. For instance, 
\citet{gfp03} examined the degree to which Doppler boosting of the 
radio and X-ray emission could produce dispersion in the $L_{\rm R}$ vs. 
$L_{\rm X}$ relationship, and concluded that the tightness of the 
correlation between $10^{34}$ and $10^{36.5}$ erg s$^{-1}$ implies that 
the bulk motion of the jets can be no more than $\sim 0.8c$. However, they
specifically omitted the data from GRS 1758--258, which they assumed was 
quenched in the radio; including that source and the neutron star LMXBs could
point toward a larger importance for Doppler boosting. 

The presence of a hard surface near the inner edge of the accretion disk 
is another factor that distinguishes neutron star LMXBs from
those containing black holes. The hard surface must arrest the accretion flow,
which would form a boundary layer \citep[e.g.,][]{ps01} that could alter 
the structure of a jet or produce photons that cool the electrons in the jet. 
However, it is difficult to believe that this is the explanation for the 
faintness of the radio emission from the neutron stars, because there are 
not significant differences in the X-ray spectral or timing components seen 
from the black hole and neutron star systems that can be attributed 
unambiguously to such a boundary layer in the latter systems 
(Fig.~\ref{fig:spec} and \ref{fig:pds}).

Finally, it has also been suggested that the depth of the gravitational 
potential out 
of which the jets are produced can greatly influence their strength 
\citep[e.g.,][]{bp82,wc95,mei01}. 
In black hole systems, the inner radius at which a stable accretion disk 
can exist depends on the black hole spin; in a neutron star systems, it is 
most likely set by either the surface of the star, or by a dipole field 
anchored to the star. The low radio luminosities of GRS 1758--298, 
\tertwo, 4U 1812--12, and SLX 1735--269 could then be explained by the 
relative shallowness of their potentials. For the neutron stars, the difference
would be due to the nature of the compact object; for GRS 1758--258, 
the difference could be explained if the black hole was either 
non-rotating, or rotating with an angular momentum opposite to that of the 
accretion disk.

\section{Conclusions}

We have obtained upper limits on the radio emission from three neutron star 
LMXBs,
concurrent with X-ray observations that demonstrated that they were in a hard 
X-ray state. We have compared these upper limits to the radio luminosities of 
several black holes in very similar hard states, and found that the neutron 
star systems are as faint as or fainter than the black hole candidate with the 
lowest observed radio luminosity, GRS 1758--258. The difference in radio 
luminosity can partly be attributed to the lower masses of the neutron star 
systems, which on theoretical and observational grounds is expected to 
decrease as $M^{0.8}$. However, there still remains a factor of 30 scatter in
the radio luminosities of black hole and neutron star X-ray binaries, 
particularly at X-ray luminosities of a few percent Eddington. We find no 
obvious differences 
in the X-ray timing and spectral properties that explain this dispersion. 

Two future observations could help resolve why there is such a large 
dispersion in the radio luminosities of black hole X-ray binaries, and why 
neutron stars tend to fall on or below the faint end of this dispersion. 
First, it is important to either detect, or place much stricter upper limits 
on,
the radio luminosities of neutron star LMXBs in the X-ray hard state. This 
would
help resolve whether there is indeed a physical difference that causes the 
relative faintness of neutron star LMXBs, or whether the three sources we 
observe just happen to fall on the low end of a dispersion in radio luminosity.
Second, radio observations of the fainter radio sources, i.e. GRS 1758--298 and
the neutron star LMXBs, at lower X-ray luminosities could reveal whether they 
are indeed quenched near $L_{\rm Edd} \sim 0.01$, or whether another parameter
is needed to explain the radio luminosity of X-ray binaries in hard states.

\acknowledgments{We are grateful to the VLA and \rxte\ planning teams,
particularly B. Clark and E. Smith, for coordinating these observations.
We also thank 
S. Markoff and R. Fender for helpful conversations
and encouragement while pursuing this work. This research has made use of data 
obtained through the High Energy Astrophysics Science Archive Research Center,
provided by NASA through the Goddard Space Flight Center. \thankhubble}

\begin{deluxetable}{lcccccc}
\tablecolumns{7}
\tablewidth{0pc}
\tablecaption{VLA Observations\label{tab:vla}}
\tablehead{
\colhead{Source} & \colhead{RA} & \colhead{DEC} & \colhead{Date} & 
\colhead{$\nu$} & \colhead{$T_{\rm Exp}$} & \colhead{$S_{\nu}$} \\
\colhead{} & \multicolumn{2}{c}{(J2000)} & \colhead{} & \colhead{(GHz)} &
 \colhead{(h)} & \colhead{(mJy)}
}
\startdata
4U 1812--12 & 18 15 06.18 & --12 05 47.1 & 2001 Jun 5 & 8.4 & 0.25 & $< 0.2$ \\
SLX 1735--269 & 17 38 17.12 & --26 59 38.6 & 2001 Nov 5 & 8.4 & 1.5 & $<0.056$ \\
	&		&	&		& 1.5 & 1.5 & $<0.16$ \\
\tertwo & 17 27 33.15 & --30 48 07.8 & 2001 Sep 25 & 8.4 & 1& $<0.025$ \\
		&		&	&		& 4.8 & 1 & $<0.035$ \\
		&		&	&		& 1.4 & 1 & $<0.41$
\enddata
\tablecomments{Positions were taken from \citet{rtb02} and \citet{wil03}.}
\end{deluxetable}

\begin{deluxetable}{lccccc}
\tablecolumns{6}
\tablewidth{0pc}
\tablecaption{\rxte\ Observations\label{tab:xte}}
\tablehead{
\colhead{Source} & \colhead{Date} & \colhead{$T_{\rm Exp}$} & \colhead{No. PCU} & \colhead{$C$} & \colhead{$HC$} \\
\colhead{}& \colhead{} & \colhead{(ks)} & \colhead{} & \colhead{(c s$^{-1}$ PCU$^{-1}$)} & \colhead{}
}
\startdata
4U 1812--12 & 2001 Jun 6 -- Jul 21 & 21 & 3.0 & 41 & 0.824 \\ 
SLX 1735--269 & 2001 Nov 5 -- 7 & 22 & 2.8 & 31 & 0.693 \\ 
\tertwo & 2001 Sep 25; Nov 9 -- 11 & 48 & 3.3 & 54 & 0.706 \\ 
\enddata
\tablecomments{For each source, we list the date of the PCA observations, 
the exposure time, the average number of active proportional counter units, 
the count rate, and the hard color.}
\end{deluxetable}

\begin{deluxetable}{lccc}
\tabletypesize{\scriptsize}
\tablecolumns{4}
\tablewidth{0pc}
\tablecaption{X-ray Spectra\label{tab:spec}}
\tablehead{
\colhead{Parameter} & \colhead{4U 1812--12} & \colhead{SLX 1735--269} & \colhead{\tertwo}
}
\startdata
$N_{\rm H}$ (cm$^{-2}$; fixed) & 1.5 & 1.5 & 1.0 \\
$kT_{\rm bb}$ (keV) & $0.59^{+0.03}_{-0.04}$ & $0.48^{+0.03}_{-0.12}$ & \nodata \\
$N_{\rm bb}$ (km [10 kpc]$^{-1}$) & $73^{+16}_{-19}$ & $150^{+1150}_{-70}$ & \nodata \\
$\Gamma$ & $1.70^{+0.03}_{-0.05}$ & $2.00^{+0.03}_{-0.03}$ & $1.96^{+0.02}_{-0.0}$ \\
$E_{\rm fold}$ (fixed) & 300 & 300 & 300 \\ 
$R$ & $0.27^{+0.07}_{-0.02}$ & $0.45^{+0.05}_{-0.09}$ & $<0.003$ \\
$\xi$ (erg cm s$^{-1}$) & 0.25 & 654 & \nodata \\
$N_{\Gamma}$ (ph cm$^{-2}$ s$^{-1}$) & $0.10^{+0.05}_{-0.02}$ & $0.11^{+0.01}_{-0.01}$ & $0.207^{+0.001}_{-0.001}$ \\
$\chi^2/\nu$ & 92/87 & 78/88 & 106/90 \\ [5pt] 
$F_{2-10 {\rm keV}}$ ($10^{-10}$ erg cm$^{-2}$ s$^{-1}$)\ & 4.1 & 3.0 & 5.6 \\
$F_{10-20 {\rm keV}}$ ($10^{-10}$ erg cm$^{-2}$ s$^{-1}$) & 2.5 & 1.8 & 2.4 \\
$F_{20-200 {\rm keV}}$ ($10^{-10}$ erg cm$^{-2}$ s$^{-1}$) & 6.8 & 1.9 & 4.9 \\
\enddata
\tablecomments{All fluxes are de-absorbed.}
\end{deluxetable}

\begin{deluxetable}{lccc}
\tabletypesize{\scriptsize}
\tablecolumns{4}
\tablewidth{0pc}
\tablecaption{X-ray Timing\label{tab:pds}}
\tablehead{
\colhead{Parameter} & \colhead{4U 1812--12} & \colhead{SLX 1735--269} & \colhead{\tertwo}
}
\startdata
constant ($\times 10^{-6}$) & 11(4) & 4(1) & 16(5) \\
$\nu_{1}$, $W_{1}$, $N_{1}$ & 0, 0.25(3), 0.039(3) 
& 0, 0.28(1), 0.056(5) 
& 0, 0.9(2), 4.7(6)$\times10^{-3}$ \\
$\nu_{2}$, $W_{2}$, $N_{2}$ & $<0.5$, 2(1), 2(1)$\times10^{-3}$
& 0, 3.0, 2.0(9)$\times10^{-3}$
& 0, 7(1), 1.5(4)$\times10^{-3}$ \\
$\nu_{3}$, $W_{3}$, $N_{3}$ & 0, 25(6), 5(2)$\times10^{-4}$
& 0, 19.3(9), 1.0(1)$\times10^{-3}$
& \nodata \\
$\nu_{4}$, $W_{4}$, $N_{4}$ & \nodata
& 0, 371(90), 4.2(7)$\times10^{-5}$ 
& \nodata \\
$\nu_{5}$, $W_{5}$, $N_{5}$ & \nodata
& 0.64(1), 0.17(7), 4.4(5)$\times10^{-3}$
& \nodata \\
$\nu_{6}$, $W_{6}$, $N_{6}$ & \nodata
& 0.62(8), 1.4(2), 6(1)$\times10^{-3}$
& \nodata \\
$\chi^2/\nu$ & 135/98 & 119/89 & 76/54 
\enddata
\tablecomments{Power spectra were fit with Lorentzians, the parameters of 
which were the centroid ($\nu_{\rm i}$), width ($W_{\rm i}$) and 
normalization ($N_{\rm i}$).}
\end{deluxetable}

\end{document}